\documentclass[prl,twocolumn]{revtex4}
\usepackage{graphicx}
\usepackage{amsmath}

\begin{document}

\title{Triggered qutrits for Quantum Communication protocols}

\author{G. Molina-Terriza$^1$, A. Vaziri$^1$, J. \v{R}eh\'{a}\v{c}ek$^2$, Z. Hradil$^2$ and A. Zeilinger$^1,^3$}

\affiliation{$^1$Institut f\"ur Experimentalphysik, Universit\"at
Wien, Boltzmanngasse, 5, A-1090, Vienna, Austria\\ $^2$ Department
of Optics, Palacky University, 17. listopadu 50, 772 00 Olomouc,
Czech Republic\\$^3$Institute for Quantum Optics and Quantum
Information, Austrian Academy of Sciences}

\begin{abstract}
A general protocol in Quantum Information and Communication relies
in the ability of producing, transmitting and reconstructing, in
general, qunits. In this letter we show for the first time the
experimental implementation of these three basic steps on a pure
state in a three dimensional space, by means of the orbital
angular momentum of the photons. The reconstruction of the qutrit
is performed with tomographic techniques and a Maximum-Likelihood
estimation method. In this way we also demonstrate that we can
perform any transformation in the three dimensional space.
\end{abstract}

\date{\today}

\maketitle

 One of the main objectives in Quantum Information
is exploring the possibilities of applying quantum systems in
communication and computation protocols. Usually, these protocols
use the information encoded in two dimensional systems, better
known as qubits. Nevertheless, some proposals show that higher
dimensional systems are better suited for certain purposes
\cite{Bourennane01a, Bechmann00a, Bechmann00b, Divincenzo00a,
Bartlett00a, Karimipour01a, Ambainis01a,Bruknercomplex03}. On a
more fundamental level, higher dimensional Hilbert spaces provide
novel counter-intuitive examples of the relationship between the
quantum and the classical information, which cannot be found in
two-dimensional systems \cite{Jozsa00}.

Encoding qunits (systems with $n$ different orthogonal states)
with photons has been experimentally demonstrated using
interferometric techniques, such as time-bin schemes
\cite{Riedmatten02QIC} and superpositions of spatial modes
\cite{Zukowski97multiport}. Up to now, the only
non-interferometric technique of encoding qunits in photons is
using their orbital angular momentum or, equivalently, their
transversal modes \cite{Mair01a,Molina-Terriza02}. Orbital
angular momentum modes usually contain dark spots which regularly
exhibit phase singularities.

The orbital angular momentum of light has already been used to
entangle  and concentrate the entanglement of two photons
\cite{Mair01a, Vaziri03a}. This entanglement has also been shown
to violate a two particle three-dimensional Bell inequality
\cite{Vaziri02b}. There have been proposals of some experimental
techniques to engineer entangled qunits in photons
\cite{Molina-Terriza02, Torres03a, Vaziri02a}. In this paper we
experimentally demonstrate all the basic steps of a higher
dimensional quantum communication protocol.

In a general communication scheme, prior to the sharing of
information, the two parties, say Alice and Bob, have to define a
procedure which will assure that the signal sent by one party is
properly received by the other one. Usually, this scheme works as
follows: First, Alice prepares a signal state she wants to send.
Bob will measure it and communicate the result to Alice, who will
correct the parameters of her sending device following Bob's
indications. This process will repeat itself until the two
parties adjust the corresponding devices. After this step is
fulfilled, Alice can rely that any subsequent signal which is
sent is properly received.

Using pairs of photons entangled in orbital angular momentum, we
can prepare any qutrit state, transmit it, and measure it. The
preparation is done by projecting one of the two photons onto
some desired state. This nonlocally projects the second photon
onto a corresponding state. This state may be transmitted to Bob
and finally measured by him. The measurement employs tomographic
reconstruction. This last step is usually a technically demanding
problem, inasmuch as it needs the implementation and control of
arbitrary transformations in the quantum system's Hilbert space.

On theoretical grounds, one convenient basis which describes the
transversal modes of a light beam fulfilling the paraxial
approximation is the Laguerre-Gaussian (LG) functions basis:
$LG_{p,m}(x,y)$. Here $m$ is the order of the phase dislocation
characteristic of this set of functions and it accounts directly
for the orbital angular momentum of the Laguerre-Gaussian mode in
units of $\hbar$\cite{Allen92a, He95a}. The other parameter $p$
is a label which is related to the number of radial nodes of the
mode and $(x,y)$ refer to any point in a plane perpendicular to
the beam propagation direction. The LG functions form a complete
and orthonormal basis for any complex function in the transveral
plane.

Holographic techniques can be used to transform $LG$ modes
\cite{Bazhenov90a}. Conveniently prepared holograms change the
phase structure of the incoming beam, adding or removing the
phase dislocations related with the orbital angular momentum.
Whereas optical single mode fibers act as a filter for all higher
$LG$ modes, i.e. only the $LG_{00}$, or Gaussian, mode can be
transmitted, the combination of holograms and single mode fibers
project the incoming photon into different states. In this way we
can define the basis of the experimentally accessible states as:
\begin{equation}
\label{base}
 \langle \vec{x}|0\rangle =LG_{0,0}(x,y),~ |m\rangle =H_m(\vec{0})|0\rangle ,
\end{equation}
where the vector $|0\rangle $ is the mode of the fiber used to
detect the photon, $\vec{x}$ is a shortcut to represent any point
in the transversal space, $m$ is a positive or negative integer,
and $H_m(\vec{0})$ is the operator which describes the action of
the $m$-th order hologram when it's centered, relative to the
fiber.

Although the mode $|m\rangle$ posses orbital angular momentum of
$m\hbar$, they are not pure $LG$-modes. However, they can be
described as coherent superpositions of different $LG$-modes with
the same $m$, but different $p$'s. In this sense, the basis we
have constructed in (\ref{base}) is not complete, since it does
not expand the $LG$ basis. In the following we refer to all modes
belonging to the subspace (\ref{base}) as ``inner'' modes and the
rest of the modes will be addressed as ``outer'' modes.

Thus, any displaced hologram and, in general, any linear operator
which acts on our Hilbert space can be expressed like
$H_m(a,b)=\sum_{i=-\infty}^{+\infty}c_i(a,b)H_i(0)+\gamma(a,b)\Gamma$
where $a,b$ are the displacements of the hologram along the
orthogonal directions in the transversal plane, relative to its
centered position. The operator $\Gamma$ accounts for the
possibility that the displaced hologram is performing
transformations between ``outer'' and ``inner'' states, i.e.
transforming any ``inner'' state into an ``outer'' one, or the
other way round.

\begin{figure}
\begin{center}
\includegraphics[width=.4\columnwidth,angle=0]{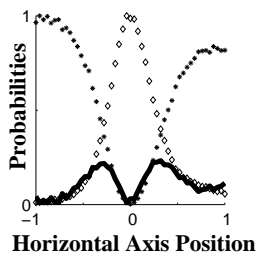}
\caption{Effect of a hologram on an initially prepared $|0\rangle
$ mode. This mode is transformed by means of a horizontally
displaced hologram of charge $m=+1$. The resulting state is
projected onto the basis states. In this case the modes
$|0\rangle$ and $|1\rangle$ are expected to contribute the most
to the transformed state. Asterisks: projection onto the
$|0\rangle$ mode. Diamonds: projection onto the $|1\rangle$ mode.
The solid line represents the projection onto ``outer'' the
modes, i.e. those not belonging to the basis (\ref{base}). This
projection is found by subtracting to the total number of events
those corresponding to the projection onto the elements of the
basis. The two positions of the hologram where the transfer to
``outer'' modes is maximum are taken as new elements of the
basis.}
\end{center}
\end{figure}
The value of $\gamma$ can be estimated experimentally. In Fig.~1
we present an example of such a measurement. It is observed how
there are two positions where the contribution of the ``outer''
modes is specially high.

 Up to now the most convenient way of transforming an
OAM state is to employ holograms. Yet, as discussed above, these
holograms might also perform unsought transformations between
``inner'' and ``outer'' modes. To avoid this problem we enlarge
our Hilbert space with some selected ``outer'' vectors.
 We choose eight different positions of the holograms as new operators
which, together with a projection into the $|0\rangle $ mode and a
Gram-Schmidt orthonormalization, enlarge our natural Hilbert
space. In the present work we enlarged the basis with four
positions for every differently charged hologram used in the
experiment. Each of this four points correspond to the two
vertical and the two horizontal positions where the probability
of transforming a state of the basis into an``outer'' mode is
bigger.

The enlargement of the basis (\ref{base}) allow us to represent
more precisely the effect of the hologram on our beam but, as a
drawback we need more measurements to estimate the state of the
photon, as there are more dimensions in our Hilbert space. This
problem, considered together with the imperfections of the
holograms and possible systematic drifts due to experimental
misalignments during the measurements, makes it a natural option
to turn to maximum likelihood (ML) schemes for reconstructing the
transformed states.

\begin{figure}
\begin{center}
\includegraphics[width=.9\columnwidth,angle=0]{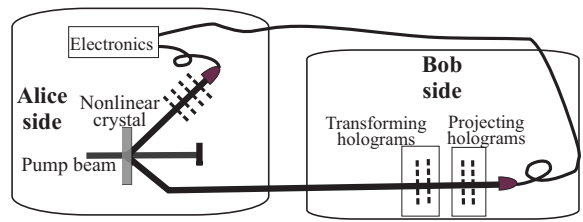}
\caption{Experimental set-up. A $351$nm wavelength laser pumps a
BBO crystal. The two generated $702$nm  down-converted photons
are send to Alice and Bob's detectors respectively. Before being
detected each photon propagates through a set of holograms. Each
photon was coupled into single mode fibers and directed to
detectors based on avalanche photo diodes operating in the photon
counting regime.}
\end{center}
\end{figure}
The experimental set-up it is shown in Fig.~2 . A $351$nm
wavelength Argon-ion laser pumps a $1.5-$mm-thick BBO
($\beta$-barium-borate) crystal cut for Type I phase matching
conditions. The crystal is positioned such as to produce
down-converted pairs of equally polarized photons at a wavelength
of $702$nm emitted at an angle of $4^\circ$ off the pump
direction. These photons are directly entangled in the orbital
angular momentum degree of freedom. Alice can manipulate one of
the downconverted photons, while the other is sent to Bob. Before
being detected, Bob's photon traverses two sets of holograms.
Each set consists of one hologram with charge $m=1$ and another
with charge $m=-1$. The first set of holograms provides the means
of a transformation in the three dimensional space expanded by
the states $|-1\rangle , |0\rangle $ and $|1\rangle $. The second
set, together with a single mode fiber and a detector, act as a
projector onto the three different basis states. All this
elements are Bob's receiving device. Alice's photon also traverses
a set of holograms, which together with the source, and the
detector on Alice side, act as Alice's sending device. Whenever
Alice detects one photon, this initiates the transmission of a
photon to Bob. By means of the quantum correlations between the
entangled photons, Alice can radically control the state of the
photon sent to Bob. In order to adjust properly their respective
devices, Bob has to perform a tomographic measurement of the
state he is receiving and classically communicate to Alice the
result.

In our experiment, the tomographic reconstruction of Bob's
received qutrit state was realized in two independent steps
trying to avoid any bias from `a priori' information. First, the
Vienna team by measuring Alice's photon projected the photons in
Bob's side and then performed the required measures. The minimum
number of measurements to reconstruct the three dimensional state
sent to Bob is $9$. This number increases to $121$ for our
enlargement to a $11$-dimensional Hilbert space. In the end, to
exploit the power of the ML reconstruction and to minimize
errors, the number of different projections was around $2400$.
The results of these measurements, together with the projecting
vectors, were sent to the Olomouc team who, without a previous
knowledge of which was the state projected by Alice,
reconstructed the density matrix describing the state of the
photon in Bob's side. As will be shown below in all the cases the
reconstructed three dimensional state was a coherent
superposition of the three ``inner'' vectors, whose relative
weights and phases could be effectively controlled demonstrating
that any qutrit state could be sent. The noise and incoherence
were within the Poissonian noise level, which is an indication of
the reliability of the tomographic measurement.

The transforming set of holograms was analyzed to properly
describe the transformation done. From the description of each
single hologram, we could express the action of each
transformation set in the following way:
\begin{equation}
\begin{split}
\label{transfo}
& \langle \vec{x_1}|H_1(a_{+1},b_{+1})H_{-1}(a_{-1},b_{-1})|\vec{x_2}\rangle = \nonumber \\
& \exp(-i\arctan(\frac{y_1-a_1}{x_1-b_1})+i
\arctan(\frac{y_1-a_{-1}}{x_1-b_{-1}}) \nonumber \\
& -i k_x x_1-i k_y y_1)\delta(x_1-x_2,y_1-y_2)
\end{split}
\end{equation}
where $k_x$ and $k_y$ are free parameters which depend on the
alignment procedure and on the holographic grating, and $a_\pm1,
b_\pm1$ represent the displacement of the two holograms. Each set
of holograms is described completely by eight parameters: the
number of maximum coincidences, the width of the beam, $4$
numbers to determine the centered position of each hologram and
the two parameters $k_x$ and $k_y$.

The estimation of these eight parameters was performed by fitting
four different experimental curves. The data which conformed the
curves were taken by sending to Bob a photon prepared in the
$|0\rangle $ state. Bob fixed one of his holograms in one
determined position and performed a scan on one of the axes of the
other hologram. The resulting state was again projected to the
$|0\rangle $ state, i.e. one of this curves can be described by
$\langle 0|H_1(x,0)H_{-1}(1,1)|0\rangle $ as a function of $x$.
Each of the four curves corresponds to the scan of all the axes of
the two holograms.

The projection measurements were made by moving the
transformation set of two holograms into around $2400$ different
positions and counting the number of coincident detections which
took place in 2 seconds. For every position, we counted typically
a few hundred of coincidences per second. The complete time for
each of this measurements was around 6 hours. After this time some
slight misalignments where detected, which could be compensated
mainly due to the large number of different projections taken and
the reconstruction process

The registered data were processed using the maximum-likelihood (ML)
reconstruction algorithm. Assuming that the statistics of
the detection events at low intensities is Poissonian,
the joint probability of observing registered data reads,
\begin{equation}\label{gabi-lik}
{\cal L}=\prod_j (N p_j)^{n_j} e^{-N p_j}/n_j!,
\end{equation}
where $N$ is the mean number of qutrits subject to each measurement
of which $n_j$ were found in the state
$|j\rangle=H_1(a^j_{+1},b^j_{+1})H_{-1}(a^j_{-1},b^j_{-1})|0\rangle$,
and $p_j=\mathrm{Tr}\{|j\rangle\langle j|\rho_\mathrm{Bob}\}$ are the corresponding
probabilities.

In accordance with the Bayes theorem \cite{Bayes-class},
$\mathcal{L}$ quantifies the likelihood of Bob's state
$\rho_\mathrm{Bob}$ in view of the measured data. The state
having highest likelihood is picked up as the result of the
reconstruction. ML estimation is known to be asymptotically
efficient \cite{fisher,Helstrom} and all existing physical
constraints such as positivity of $\rho$ can easily be
incorporated into the reconstruction process.

From the technical point of view the maximum of functional
\eqref{gabi-lik} is found by iterating the extremal equation,
\cite{Rehacek01} $R\rho R=G\rho G$ starting from the maximally
mixed state. Hermitian operators
$R=\sum_j(n_j/p_j)|j\rangle\langle j|$ and $G=(\sum_j
n_j)/(\sum_j p_j)\sum_j|j\rangle\langle j|$ are functions of the
measurements and detected data.

Let us mention that though the reconstruction is done on the full
$3+8$ dimensional Hilbert space spanned by the ``inner'' and
``outer'' states, we are interested only in the ``inner''
subspace. Therefore all the reconstructed states are projected to
this subspace to simplify the discussion.

\begin{figure}
\includegraphics[width=.9\columnwidth]{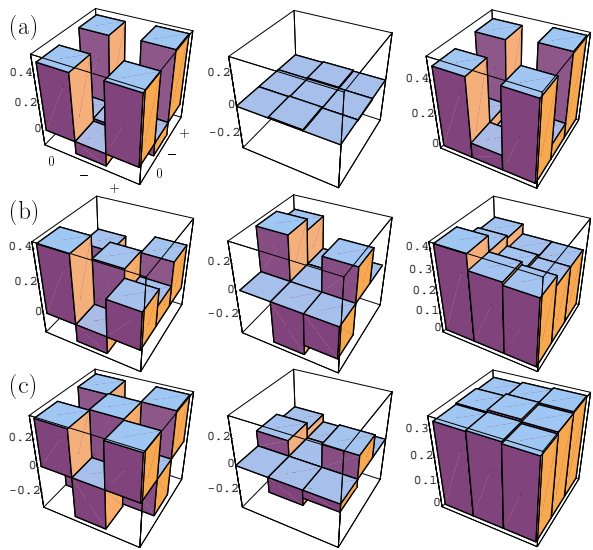}
\caption{Results of quantum state tomography applied to three
different remotely prepared states of Bob's qutrits:(a)
$0.68|0\rangle+0.71|1\rangle-0.14|\text{--1}\rangle$; (b)
$0.65|0\rangle+0.53\exp(-i0.26\pi)|1\rangle+
0.55\exp(-i0.6\pi)|\text{--1}\rangle$; (c)
$0.58|0\rangle+0.58\exp(-i.05\pi)|1\rangle+
0.58\exp(-i.89\pi)|\text{--1}\rangle$. Left and middle panels
show real and imaginary parts of the reconstructed density
matrices; right panels visualize the absolute values of those
elements for better comparison of how large are the contributions
of the three basic states. From the results it is shown that
Alice can control both the relative amplitudes and phases of the
sent states.
\label{fig:gabi-results}}
\end{figure}

As we have already explained, different projections done by Alice
translate through the entanglement into different state
preparations on Bob's side. Three such remote preparations are
shown in Fig.~\ref{fig:gabi-results}.
All of them were found to be very nearly pure states, their
largest eigenvalues and corresponding eigenvectors being (a)
$\lambda_{max}=0.99$,
$|e_{max}\rangle=0.68|0\rangle+0.71|1\rangle-0.14|-1\rangle$; (b)
$\lambda_{max}=0.99$,
$|e_{max}\rangle=0.65|0\rangle+0.53\exp(-i0.26\pi)|1\rangle+
0.55\exp(-i0.6\pi)|-1\rangle$;  (c) $\lambda_{max}=0.99$,
$|e_{max}\rangle=0.58|0\rangle+0.58\exp(-i0.05\pi)|1\rangle+
0.58\exp(-i0.89\pi)|-1\rangle$. In case (a) Alice tried to prepare
an equal-weight superposition of $|0\rangle$ and $|-1\rangle$
basis states. Utilizing the conservation of the orbital momentum
in downconversion, this was easily done by projecting her qutrit
along the ray $|0\rangle+|1\rangle$: Her hologram with the
positive charge  was taken out of the beam path and the center of
the other one was displaced with respect to the beam by a
determined translation vector. Cases (b) and (c) both represent an
equal-weight superposition of the three states, but with
different relative phases, showing that besides the relative
intensities, we could also control the relative phases. Other
qutrits reconstructed (not shown in Fig.~3), showed an effective
suppression of the $|0\rangle$ mode, through destructive
interference from the two holograms. The result was
$\lambda_{max}=0.97$,
$|e_{max}\rangle=0.26|0\rangle+0.68\exp(i0.11\pi)|1\rangle+
0.68\exp(-i0.21\pi)|-1\rangle$.

As can be deduced from the maximum eigenvalue of all the data,
the purity of the reconstructed states was over $97\%$. On the
other hand, by direct comparison of the measured data and the
data estimated by the reconstructed matrix, the error was
comparable to the statistical Poissonian noise, which
demonstrates the reliability of the tomography.

In conclusion, here we have demonstrated the feasibility of a
point to point communication protocol in a three-dimensional
alphabet. Using the orbital angular momentum of photons, we have
implemented three basic tasks inherent in any communication or
computing protocol: preparation, transmission and reconstruction
of a qutrit. In particular, the reconstruction was exercised with
a tomographic estimation of the density matrix, which also
demonstrates that we could perform any rotation of the states.

This work was supported by the Austrian Science Foundation (FWF),
project number SFB 015 P06, by project LN00A015 of the Czech Ministry of
Education, and by the European Commission,
contract no. IST-2001-38864, RAMBOQ. G.M.-T. is a Marie Curie
Fellowship.


\begin{thebibliography}{10}
\bibliographystyle{unsrt}

\bibitem{Bourennane01a}
M.~Bourennane, A.~Karlsson, and G.~Bj{\"o}rk, Phys. Rev. A {\bf
64}, 012306 (2001).

\bibitem{Bechmann00a}
H.~Bechmann-Pasquinucci and A.~Peres, Phys. Rev. Lett. {\bf 85},
3313 (2000).

\bibitem{Bechmann00b}
H.~Bechmann-Pasquinucci and W.~Tittel, Phys. Rev. A {\bf 61},
62308  (2000).

\bibitem{Divincenzo00a}
D.~P. DiVincenzo, T.~Mor, P.~W. Shor, J.~A. Smolin, and B.~M.
Terhal, Comm. Math. Phys. {\bf 238}, 379 (2003).


\bibitem{Bartlett00a}
S.~D. Bartlett, H.~de~Guise, and B.~C. Sanders, Proceedings of
IQC'01, 344 (2001). quant-ph/0011080.

\bibitem{Karimipour01a}
V.~Karimipour, S.~Bagherinezhad, and A.~Bahraminasab, Phys. Rev. A
{\bf 65}, 042320 (2002).

\bibitem{Ambainis01a}
Andris Ambainis, Proceedings of STOC, pages 134--142 (2001).

\bibitem{Bruknercomplex03}
C.~Brukner, M.~Zukowski and A.~Zeilinger, Phys. Rev. Lett. {\bf
89}, 197901 (2002).

\bibitem{Jozsa00}
R.~Jozsa, and J.~Schlienz, Phys. Rev. A {\bf 62}, 012301 (2000).

\bibitem{Riedmatten02QIC} H.~de~Riedmatten, I. Marcikic, H.
Zbinden and N. Gisin, Quantum Information and Computation {\bf 2},
425 (2002).

\bibitem{Zukowski97multiport} M. Zukowski, A. Zeilinger, M. A.
Horne, Phys. Rev. A {\bf 55}, 2564 (1997).

\bibitem{Mair01a}
A.~Mair, A.~Vaziri, G.~Weihs, and A.~Zeilinger, Nature {\bf 412},
313 (2001).

\bibitem{Molina-Terriza02} G.~Molina-Terriza, J.~P.~Torres and
L.~Torner, Phys. Rev. Lett. {\bf 88}, 013601 (2002).

\bibitem{Vaziri03a}
A.~Vaziri, J. W. Pan, T. Jennewein, G. Weihs and A. Zeilinger.
 \newblock  quant-ph/0303003.

\bibitem{Vaziri02b}
A.~Vaziri, G.~Weihs and A.~Zeilinger, Phys. Rev. Lett., {\bf 89},
240401 (2002).

\bibitem{Vaziri02a}
A.~Vaziri, G.~Weihs, and A.~Zeilinger, J. Opt. B: Quantum
Semiclass. {\bf 4}, S47--S50 (2002).

\bibitem{Torres03a}
J.~P.~Torres, Y.~Deyanova, L.~Torner and G.~Molina-Terriza, Phys.
Rev. A {\bf 67}, 052313 (2003). J.~P.~Torres, A.~Alexandrescu and
L.~Torner, quant-ph/0306105.

\bibitem{Allen92a}
L.~Allen, M.~W. Beijersbergen, R.~J.~C. Spreeuw, and J.~P.
Woerdman, Phys. Rev. A {\bf 45}, 8185 (1992).

\bibitem{He95a}
H.~He, M.E. Friese, N.R. Heckenberg, and H.~Rubinsztein-Dunlop,
Phys. Rev. Lett. {\bf 75}, 826 (1995).

\bibitem{Bazhenov90a}
V.~Yu Bazhenov, M.~V. Vasnetsov, and M.~S. Soskin, JETP Lett.
{\bf 52}, 429 (1990). N.~R.~Heckenberg, R.~McDu., C.~P.~Smith,
and A.~G.~White., Opt. Lett. {\bf 17}, 221 (1992).

\bibitem{Bayes-class}
J. M. Bernardo and A. F. M. Smith, ``Bayesian Theory'' (Wiley, Chichester, 1994).

\bibitem{fisher} R.A. Fisher, Proc. Camb. Phi. Soc. {\textbf 22},
700 (1925).

\bibitem{Helstrom}  C. W. Helstrom, {\em Quantum Detection and Estimation
Theory,} (Academic Press, New York 1976).

\bibitem{Rehacek01}
J.~\v{R}eh\'{a}\v{c}ek, Z.~Hradil, and  M.~Je\v{z}ek,
 Phys. Rev. A \textbf{63},  040303(R) (2001).

\end{thebibliography}
\end{document}